\documentclass[11pt,twocolumn,twoside]{osajnl}
\usepackage{mathtools}
\usepackage{amsmath}
\usepackage{units}
\usepackage{upgreek}
\graphicspath{ {./image/} }

\journal{ao} 

\setboolean{shortarticle}{true}

\title{Phase retrieval algorithm applied to high-energy ultrafast lasers}
\author[1,2,*]{Jikai Wang}
\author[2]{Abdolnaser Ghazagh}
\author[1]{Sonam Smitha Ravi}
\author[3]{Stefan Baumbach}
\author[3]{Benjamin Dannecker}
\author[3]{Michael Scharun}
\author[3]{Dominik Bauer}
\author[2,4]{Stefan Nolte}
\author[1]{Daniel Flamm}
\affil[1]{TRUMPF Laser-und Systemtechnik GmbH, Johann-Maus-Strasse 2, 71254 Ditzingen, Germany}
\affil[2]{Friedrich Schiller University Jena, Institute of Applied Physics, Abbe Center of Photonics, Albert-Einstrein-Str.~15, 07745 Jena, Germany }
\affil[3]{TRUMPF Laser GmbH, Aichhalder Str.~39, 78713 Schramberg, Germany}
\affil[4]{Fraunhofer Institute for Applied Optics and Precision Engineering, Albert-Einstein-Str.~7, 07745 Jena, Germany}
\affil[*]{jikai.wang@trumpf.com}

\ociscodes{(140.3295) Laser beam characterization, (140.3300) Laser beam shaping, (100.5070) Phase retrieval, (320.7090) Ultrafast lasers, (140.3390) Laser materials processing.}
\doi{\url{https://doi.org/10.1364/AO.516820}}

\begin{abstract}
  A standardized phase retrieval algorithm is presented and applied to an industry-grade high-energy ultrashort pulsed laser to uncover its spatial phase distribution. We describe in detail how to modify the well-known algorithm in order to characterize particularly strong light sources from intensity measurements only. With complete information about the optical field of the unknown light source at hand, virtual back propagation can reveal weak points in the light path such as apertures or damaged components. 
\end{abstract}

\setboolean{displaycopyright}{false}

\begin{document}

\maketitle
J.~Wang \textit{et al.}, Appl.~Opt. \textbf{63}, 2518 (2024).\vspace{0.1cm }\\ 
\textbf{© 2024 Optica Publishing Group. One print or electronic copy may be made for personal use only. Systematic reproduction and distribution, duplication of any material in this paper for a fee or for commercial purposes, or modifications of the content of this paper are prohibited.}

\section{Introduction}\label{section:1}

There is constant news about extreme power and energy records for ultrafast laser systems in various amplifier architectures \cite{saraceno2019amazing, stark20211, dominik2022thin}. Even considering industrial availability, in the near future, $\sim\unit[100]{mJ}$-class lasers operating in the multi-kilowatt-regime providing subpicosecond pulses will be available \cite{dominik2022thin}. One of the main applications of these light sources is high-intensity laser-matter interaction for the efficient generation of short wavelength electromagnetic radiation or particle acceleration \cite{kneip2010bright, albert2014laser, albert20212020}. These secondary sources typically require highest peak intensities provided by the primary laser starting from $\sim \unit[1\text{E}14]{W/cm^2}$ for the generation of high-harmonics \cite{chang1997generation} up to $>\unit[1\text{E}19]{W/cm^2}$ for laser wakefield acceleration \cite{cole2015laser}. Today's industrial ultrafast lasers already provide $>\unit[1\text{E}16]{W/cm^2}$ (e.g.~at a pulse energy of $E_\text{P} = \unit[1]{mJ}$, a pulse duration of $t_\text{P} = \unit[1]{ps}$ and a focusing with numerical aperture of $\text{NA}=0.4$). The nonlinear high intensity laser-matter interaction as mentioned above, is also at hand when ultrafast lasers are used as subtle tools for micromachining \cite{flamm2021structured}. However, the new record lasers providing extreme intensities can only be profitably used for materials processing, if structured light concepts are applied to distribute the high energies into large volumes or onto large surfaces of the workpieces \cite{flamm2021structured}. Prime examples to these two strategies are cutting of glasses with non-diffracting beams \cite{jenne2020facilitated} or surface texturing with interference patterns \cite{ranke2022high}. Here, the processing optics represent the key to completely use the laser's power performance by throughput scaling. Hence, through the optical head, micromachining ($\sim\upmu$m-scale) is enabled on macroscopic dimensions ($>$ mm-scale) \cite{flamm2022multi}.

In any of the cases outlined, the generation of secondary radiation or large-scale micromachining, the requirements for the processing optics are high. Here, one primary challenge in designing the optics is often to operate below the damage thresholds of the components used and in particular those of the coatings \cite{Willemsen2022}. This example shows the importance of a thorough optical design that must be able to make predictions about the optical field at any point of the light path. The starting point for these design tasks is always a representative model of the light source. For the purpose of laser beam characterization, the standardized measurement procedure for determining the ``quality'' of the laser source via the beam propagation ratio or $M^2$ parameter \cite{siegman1998maybe} is typically used, see ISO 11146-1/2/3 \cite{ISO11146}. The use of a standardized procedure ensures the comparability of beam qualities of different laser architectures \cite{russbueldt2014innoslab, mafi2005beam, nagel2021thin}. Comparability, on the other hand, comes at the expense of describing a complex optical field with a few representative parameters only ($M^2$ factor, intrinsic astigmatism and twist parameter) \cite{ISO11146, schmidt2011real}. This, of course, results in a loss of information which is essential for the design of optical concepts. For example, a typical specification for a ``near-diffraction-limited'' beam quality for a high-power ultrafast laser source is $M^2 \lesssim 1.3$ \cite{stark20211, dominik2022thin}. The reasons for a beam quality degradation in this marginal range can nevertheless be manifold and may, for example, consist of low power amounts of spatial modes of higher order \cite{eidam2011experimental}, wavefront aberrations \cite{bueno2009temporal} or amplitude perturbations \cite{russbueldt2014innoslab}. Equally manifold will be the impact of the corresponding optical fields to local intensities within the optics or onto the target. Therefore, there is a great interest in a standardized measurement method that reveals the underlying optical field.

In this work, we combine the advantages of a measurement procedure for determining the beam quality from intensity profiles (according to ISO 11146 \cite{ISO11146}) with a method for retrieving the phase of an unknown laser beam proposed by Allen and Oxley \cite{allen2001phase}, see Sec.~\ref{section:2}. We use the measurement specifications for recording the caustics and standardized requirements regarding detectors, signal processing, etc.~\cite{ISO11146} and, thus, ensure a high degree of robustness and reproducibility in the phase reconstruction of the test laser. Besides special components for attenuating the extreme laser radiation, the required metrology consists only of the hardware that is usually available for caustics measurements anyway \cite{ISO11146}.

We apply the phase retrieval metrology to synthetic (Sec.~\ref{section:3}) and experimental data (Sec.~\ref{section:4}) from different ultrafast lasers and consider the radiation to be composed of a single optical field (complete coherence; for partial coherence see Ref.~\cite{pang2020focal}). Corresponding pulse durations are longer than $\unit[100]{fs}$ so that a description by a single wavelength (quasi-monochromatic case) is valid in a good approximation.

The versatility of our approach is illustrated by mimicking situations that are highly relevant to laser technologists, such as clipping at apertures \cite{martinez1993second} or smallest obscurations in the light path \cite{zhao2012aberration}. Knowledge about the corresponding optical field enables virtual back propagation through the optics and identifying weak points such as misaligned or degraded components.

\section{Fundamentals} \label{section:2}
Based on the study of Allen and Oxley we follow their suggestion of using an ``iterative approach'' to retrieve light's phase distribution, see Sec.\,2.3 in Ref.\,\cite{allen2001phase}, which is based on the fundamental concept of Gerchberg and Saxton \cite{gerchberg1972practical}. This is mainly due to our demand for retrieving such phase distributions that may exhibit phase jumps or even phase singularities. As we will see later, phase discontinuities are not rare when considering ultrafast lasers operating at the spatial coherence limit. When it comes to a metrology intended to be used in an industrial environment, robustness is also crucial. In this respect, we also refer to Ref.\,\cite{allen2001phase}, where the iterative method is shown to have clear advantages over techniques based on the transport-of-intensity equation \cite{merx2020beam}.

Recovering the phase distribution $\phi\left(x,y\right)$ of an unknown (transverse) optical field $E\left(x,y\right) = A\left(x,y\right) \exp{\left[\imath \phi\left(x,y\right)\right]}$ from intensity signals only $I\left(x,y\right) = \left|A\left(x,y\right)\right|^2$  has been state of the art since the fundamental studies by Fienup \textit{et al.~}\cite{fienup1986phase} at the latest---in different variants \cite{moulanier2023fast} and commercial solutions for laser beam characterization \cite{varkentina2017new}. The basis for this is the use of Fourier operators to wave optically predict light's propagation. A whole zoo of algorithms has grown out of this to design beam-shaping elements in which field quantities are exchanged iteratively for example when focused by a lens \cite{wyrowski1988iterative, ripoll2004review}. In the problem at hand, there is a whole set of intensity signals available, such as those recorded during a caustic measurement $I_j\left(x,y\right) \coloneqq I\left(x,y;z_j\right)$. We aim for retrieving the phase distribution of the optical field in a specific distance $z$ and use the angular spectrum method to determine its free space propagation by
\begin{gather} \label{eq:masterprop}
    E\left(x,y;z\right) = \mathcal{F}^{-1}\left[\mathcal{F}\left[E\left(x,y;0\right)\right] \exp\left(\imath k_z z \right) \right],
\end{gather}
where $k_z\left(k_x, k_y\right) = \sqrt{4\uppi^2/\lambda^2 - k_x^2 - k_y^2}$ with the wavevector $\mathbf{k} = \left(k_x, k_y, k_z\right)$, and $\mathcal{F}$ and $\mathcal{F}^{-1}$ as the Fourier transform and its inversion, respectively \cite{goodman2005introduction}.

\begin{figure*}
    \centering
    \includegraphics[width=1.0\textwidth]{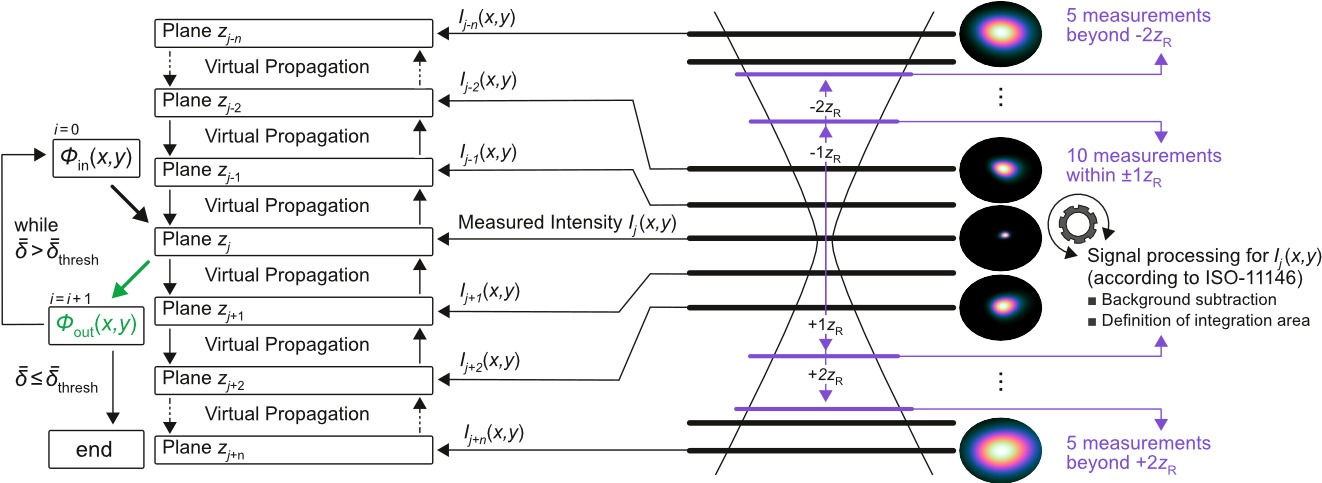}
\caption {Schematic of the phase retrieval based on an iterative Fourier transform algorithm \cite{allen2001phase}. Starting with an initial phase at propagation plane $z_j$, for example $\phi_\text{in}\left(x,y\right) = 0$, the optical field is completed using measured intensities $I_j\left(x,y\right)$. According to the ISO standard for laser beam characterization measured intensity profiles are recorded at well defined positions $z_j$ relative to the beam's Rayleigh length $z_\text{R}$ \cite{ISO11146}. After each iteration $i$, $\phi_\text{in}$ is replaced by $\phi_\text{out}$ until a termination criterion is reached, for example $\bar{\delta} \leq \bar{\delta}_{\text{thresh}} = 0.2$, cf.~Eq.~(\ref{eq:relative error mean}). }
    \label{fig:algrithm}
\end{figure*}

Starting with an initial guess for the phase distribution $\phi_\text{in}\left(x,y\right)$ at propagation step $z=z_j$, the optical field is completed using the respective measured intensity $A_j\left(x,y\right)=\sqrt{I_j\left(x,y\right)}$, see schematic in Fig.~\ref{fig:algrithm}. 
Now, the propagation to the next $z$-position is virtually calculated with the Fourier operator, cf.~Eq.~(\ref{eq:masterprop}). By iteratively inserting the measured intensity distribution, the virtual optical field will approximate the experimentally present field. After calling this sequence several times (number of iterations $i$), the method provides a phase distribution $\phi_\text{out}\left(x,y\right)$ when a certain quality measure on the ``similarity'' of the reconstructed and measured intensity profiles is reached \cite{allen2001phase}. In our case, termination is reached when the relative power reconstruction error $\bar{\delta}$ is smaller than $\bar{\delta}_{\text{thresh}} = 0.2$, see definitions in Sec.~\ref{section:3}\,\ref{section:3a}.  As denoted in the schematic of Fig.~\ref{fig:algrithm}, we propose to record the set of intensity profiles at at least 10 $z$-positions within one Rayleigh length $z_{\text{R}}$ on either side of the beam waist and beyond $2z_{\text{R}}$ from the waist \cite{ISO11146}. Known from the ISO standard, this ensures to record features from the beam's near- and far-field with sufficient sampling. Also proposed in the standard and adopted here are techniques for defining a relevant region-of-interest and for background processing \cite{ISO11146}. It has proven sufficient to perform the so-called ``coarse correction,'' cf.~ISO 11146-3 \cite{ISO11146}, in which an averaged background map is subtracted from the signal. However, when applying commercial tools for laser beam characterization, as we will also do in Sec.~\ref{section:4}\,\ref{sec:section4d}, advanced noise reduction algorithms that conform to the ISO standard are typically used anyway.

The use of commercial tools for laser beam characterization seems tailor-made to apply the proposed method. Besides the determination of widely used laser parameters like $M^2$, additionally, an unknown phase profile can be reconstructed by applying simple Fourier methods, cf.~Eq.~(\ref{eq:masterprop}), to the recorded data. 

\section{Retrieving synthetic phase profiles} \label{section:3}

In this section, we mimic distorted optical fields by adding specific aberrations based on Zernike polynomials to a pure Gaussian beam with an ideal TEM$_{00}$ profile, see Sec.\,\ref{section:3}\,\ref{section:3a}. We extend these simulations to a large number of samples and present a statistical analysis on the robustness of our method in Sec.\,\ref{section:3}\,\ref{section:montecarlo}. Through this Monte Carlo method, fundamental limitations of the procedure are discussed. In a last virtual experiment, focus profiles of an amplitude-modulated beam are analyzed. The reconstructed optical field enables virtual back propagation along the optical path to identify potential weak points in beam delivery, such as unfavorable apertures or damaged mirrors, see Sec.\,\ref{section:3}\,\ref{section:3b}.

\subsection{Wavefront aberrations from Zernike polynomials} \label{section:3a}
Wavefront aberrations are often specified in terms of a set of Zernike polynomials \cite{noll1976zernike}. Even complex wavefront distributions $W\left(x, y\right)$ can be described sufficiently with a few Zernike mode coefficients only. In the case of complete coherence, as assumed in this study, a single phase distribution $\Phi\left(x,y\right)$ can be assigned to the laser radiation. Here, wavefront and phase distributions are related by a simple normalization $\Phi\left(x,y\right) = \left(2\uppi / \lambda\right)W\left(x, y\right)$ \cite{ISO15367}. Please note that although $W\left(x,y\right)$ is defined as a continuous surface, one cannot always be found \cite{ISO15367}. The spatial dependence of a linearly polarized (scalar) optical field $E\left(x,y\right)$ can be described as follows \cite{schulze2013reconstruction}
\begin{equation}\label{eq:FeldZern}
    E\left(x,y\right) = A\left(x,y\right)\exp{\left[\imath \uppi \sum_{mn}c_{mn} Z_{mn}\left(x,y\right) \right]},
\end{equation}
with amplitude distribution $A\left(x,y\right)$ and a set of Zernike modes $\left\{Z_{mn}\left(x, y\right)\right\}$ with $n$-th radial and $m$-th azimuthal order, indexed according to Noll \cite{noll1976zernike}. Typically, access to the amplitude of the optical field is obtained via simple intensity measurements $I\left(x,y\right) = \left|A\left(x,y\right)\right|^2$. Thus, complex optical fields can be completely described with knowledge about a few mode coefficients $c_{mn}$. 

\begin{figure*}[h]
    \centering    \includegraphics[width=0.85\textwidth]{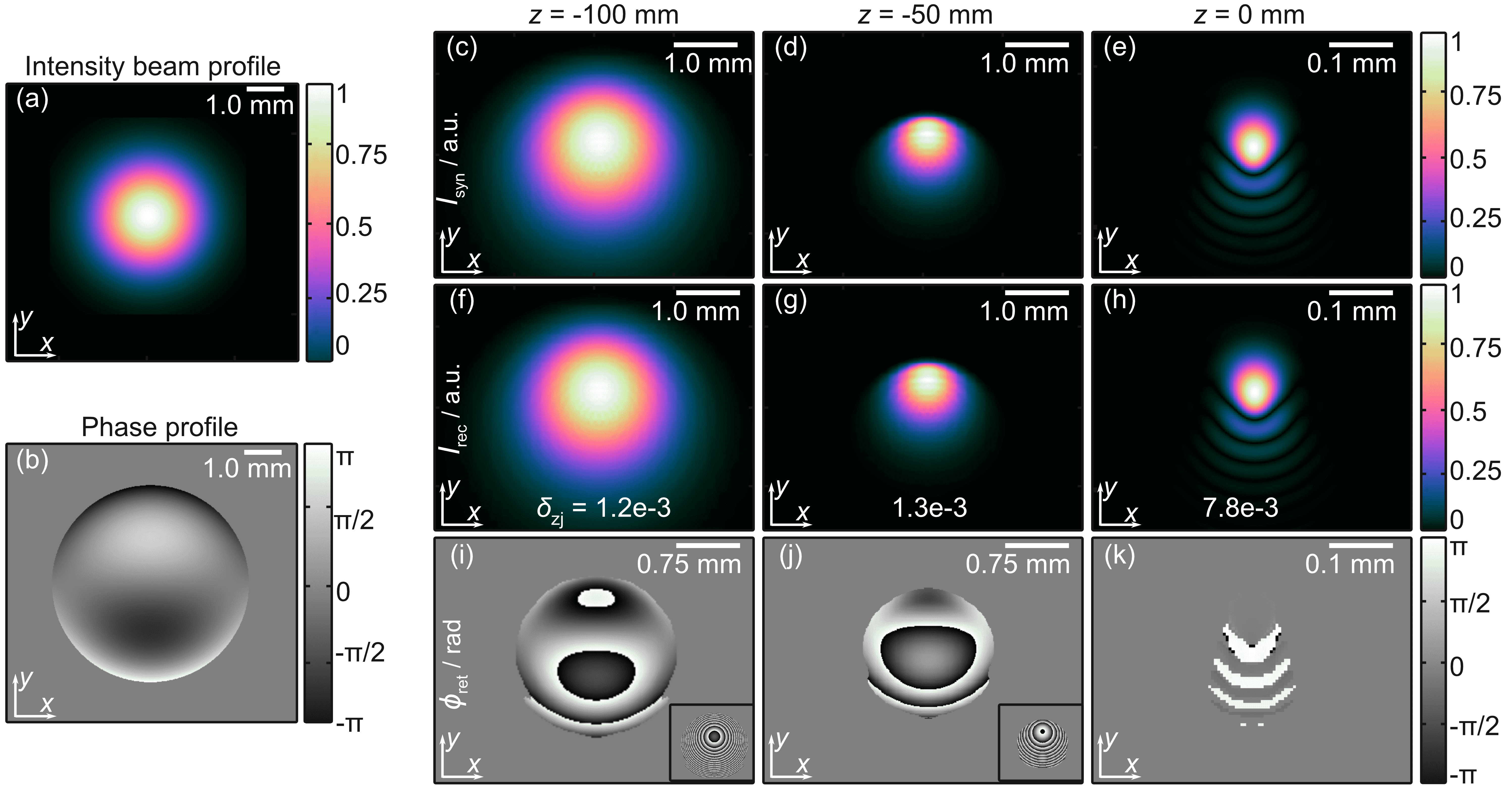}
    \caption {Phase retrieval result (synthetic data): Input intensity (a) and phase profile (b) with examples of generated intensity profiles $I_{\text{syn}}$ at three different $z$-positions at $z = \unit[-100]{mm}$ (c), $z = \unit[-50]{mm}$ (d), and in the focus at $z = \unit[0]{mm}$ (e). The corresponding retrieved intensity $I_{\text{rec}}$ and phase distributions $\phi_{\text{ret}}$ are depicted in subfigures (f) -- (k). Relative reconstruction errors $\delta_{z_j}$, cf.~Eq.~(\ref{eq:relative error}), amount to $<1\,\%$.}
    \label{fig:phasemaske1}
\end{figure*}

In our virtual experiment, we apply vertical coma $c_{\left(-1,3\right)} = 1$ (corresponds to a $1\uplambda$ peak-to-valley (PV) phase modulation) to a fundamental Gaussian beam of diameter $d_0 = \unit[5]{mm}$ and wavelength $\lambda = \unit[1030]{nm}$, see Fig.~\ref{fig:phasemaske1}\,(a) and (b). Using an ideal lens ($f=\unit[300]{mm}$) a set of $20$ beam profiles is recorded in well-defined distances with respect to the local waist characterized by $z_\text{R} \approx \unit[5]{mm}$ (conform to the ISO standard at $z_j = \left[-100, -50, -30, -20, -15\right]$\,mm, $z_j = \left[-5, -4, \dots 4, 5\right]$\,mm), $z_j = \left[15, 20, 30, 50, 100 \right]$\,mm), cf.~Fig.~\ref{fig:algrithm}.
Parts of this synthetic caustics $I_j\left(x,y\right)$ are depicted in subfigures (c), (d), and (e) of Fig.~\ref{fig:phasemaske1}. Applying the described phase retrieval to this data allows to reconstruct the optical field in amplitude and phase, see (f) -- (h) and (i) -- (k), respectively. To evaluate the quality of the reconstructions $I_{\text{rec}}$, we define the relative power reconstruction error
\begin{gather}
 \label{eq:relative error}          \delta_{z_j}=\frac{\iint_{}^{}\left|I_{\text{syn}}\left(x,y;z_j\right)- I_{\text{rec}}\left(x,y;z_j\right)\right|\text{d}x\text{d}y}{\iint_{}^{}I_{\text{syn}}\left(x,y;z_j\right)\text{d}x\text{d}y},
\end{gather}
and the $z$-averaged version
\begin{gather}\label{eq:relative error mean}
    \bar{\delta} = \sum_{j=1}^{n}\delta_{z_j}\,/n.
\end{gather}
Already after less than $10$ iterations and a computational effort of a few seconds (standard PC) we receive reconstructions with $\bar{\delta} < 1\,\%$. Another special feature of this approach becomes clear when looking at the reconstructed phase at $z = 0$ (corresponds to focus position), see Fig.~\ref{fig:phasemaske1}\,(k). Here, as known from coma-aberrated foci, the side lobes are alternately out of phase by $\uppi$. Thus, the method to iteratively retrieve phase distributions even allows to analyse optical fields with discontinuities---a clear advantage in comparison to any type of wavefront detection method providing continuous surfaces only \cite{neubert2001problem, neubert2005influences}.
Additionally, this fundamental synthetic example reveals that amplitude and phase modulations cannot be considered independently. Modulating the phase in the near-field will always vary both amplitude and phase distribution in the corresponding far-field. The phase-retrieval-based metrology is capable to detect modulations in the complex amplitude of the optical field, see Fig.~\ref{fig:phasemaske1}. Here, and in all following cases (synthetic and experimental, cf.~Sec.~\ref{section:4}), for reasons of clarity, phase profiles are set to zero at positions of weak intensity signals $I_{j, \text{thresh}}\left(x,y\right) = 0.001 \max\left[I_j\left(x,y\right)\right]$.

\subsection{Tolerance study}\label{section:montecarlo}
In the previous sections, we have identified remarkable features of the iterative phase retrieval method. However, statements about the quality of the retrieved phase distributions are still pending. Therefore, in this section, we set up an adapted Monte Carlo method statistically evaluating the robustness of our approach. This phase retrieval's input is the set of measured intensities $I_j\left(x,y\right)$, cf.~Sec.~\ref{section:2}. For evaluating the phase retrieval all influences need to be considered, which affect the intensity $I$ itself or its spatial allocation $\left(x,y;z_j\right)$. Considering the metrology at hand, we identified the transverse camera coordinates as well as the detector's noise as critical parameters. The relative position of the detector to the beam's optical axis $\Delta x$, $\Delta y$ was set to be in the range of $\unit[32]{\upmu \text{m}}$ for both transverse axes. Both limits have been determined experimentally for the employed axis stage. The upper/lower SNR limit was chosen from the noise specification of the employed camera, cf. ~Sec.~\ref{section:4}\,\ref{section4a}. Regarding the allowed aberrations, we consider a random composition of the first $12$ Zernike modes (excluding piston, tip, tilt and defocus) modifying the optical field according to Eq.~(\ref{eq:FeldZern}). Limits for mode weights $c_{mn}$ are chosen that resulting beam quality is $M^2 \lesssim 1.5$. Thus, we restrict ourselves to the beam quality range that is relevant for industry-grade ultrashort pulsed lasers \cite{russbueldt2014innoslab,mans2021high, dominik2022thin}. An overview of the Monte Carlo simulation parameters is provided in Table \ref{tab:MCparam}.
\begin{table}[]
\centering
\begin{tabular}{cc}
    \toprule
     Parameter& Lower / upper limit   \\
    \midrule
    Detector SNR (dB) & $40\dots 15$  \\
    Detector shift $\Delta x$, $\Delta y$ ($\upmu \text{m}$) & $-16 \dots 16$  \\
    Zernike weights $c_{mn}$ ($\uplambda$) & $-3 \dots 3$ \\
    \bottomrule
\end{tabular}\par
\caption{Parameter limits for Monte Carlo Simulations. For the wavefront aberrations a set of the first twelve Zernike modes are considered.}
\label{tab:MCparam}
\end{table}

Conducting the Monte Carlo simulation with synthetic data allows to compare the retrieved optical field in amplitude and phase with those that originally compose the beam under test. Regarding the amplitude reconstruction the quality parameter is already given, see $\delta_{z_j}$, $\bar{\delta}$ in Eqs.~(\ref{eq:relative error}) and (\ref{eq:relative error mean}). Considering the quality of the retrieved phase $\phi_{\text{ret}}$ we define the root-mean-square-error $\Phi_{\text{RMSE},\, z_j}$ as quality measure for the retrieved phase at a single $z$-position
\begin{multline}
\Phi_{\text{RMSE},\, z_j} = \\ \sqrt{\frac{\int_{x_1}^{x_2}\int_{y_1}^{y_2} \left[\phi_{\text{ret}}(x,y;z_j) - \phi_{\text{syn}}(x,y;z_j) \right]^2 \,\text{d}x \text{d}y}{\left(x_2-x_1\right)\left(y_2-y_1\right)}}
\end{multline}
and the $z$-position averaged version
\begin{equation} \label{eq:rmsebar}
\bar{\Phi}_{\text{RMSE}} =  \left(\sum_{j=1}^{n}\Phi_{\text{RMSE},\, z_j}/n\right) \times \left(\lambda / 2\uppi\right) 
\end{equation}
to characterize the phase reconstruction along the entire caustics at $n$ measurement positions.

The employed iterative approach to retrieve phase distributions yields $\phi\left(x,y\right)$ in a modulo-$2\uppi$ representation, see Fig.~\ref{fig:phasemaske1}\,(k). However, computing differences close to non-continuous phase signals---close to $2\uppi$ jumps---might result in large differences although the optical impact is neglectable. To compensate for this effect, all phase signals have been unwrapped before computing $\Phi_{\text{RMSE},\, z_j}$ \cite{maier2015robust}.

\begin{figure}[t]
\includegraphics[width=\columnwidth]{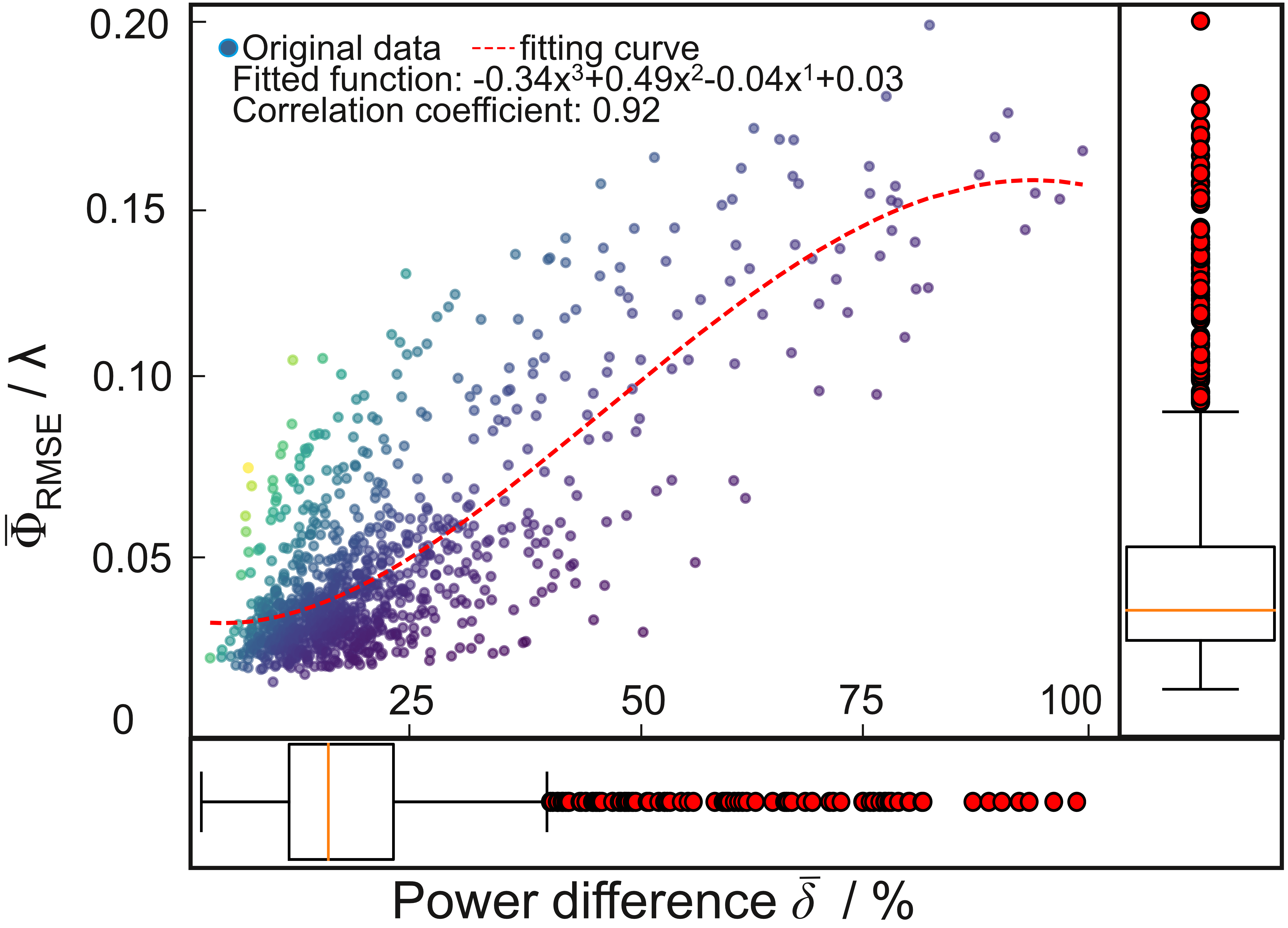}
\caption{Results of the Monte Carlo-based tolerancing. In addition to the box plots for $\bar{\Phi}_{\text{RMSE}}$ and $\bar{\delta}$ \cite{nuzzo2016box}, see right hand side and bottom, respectively, the dependency of $\bar{\Phi}_{\text{RMSE}}\left(\bar{\delta}\right)$ is shown allowing to estimate the quality of the retrieved phase from experimental, intensity-only data \cite{merx2020beam}.}
\label{fig:power vs rmse}
\end{figure}

Results of the Monte Carlo tolerancing are shown in Fig.~\ref{fig:power vs rmse} where $\bar{\Phi}_{\text{RMSE}}$ and $\bar{\delta}$ is calculated for 1000 synthetic samples within the parameter range defined in Table \ref{tab:MCparam}. As $\bar{\delta}$ converges well already after $10$ iterations, the maximum number of iterations was set to $i_{\text{max}} = 10$ for all samples. A constant number of iterations also ensures comparability of the retrieved optical fields. The $\bar{\Phi}_{\text{RMSE}}$ box plot shows a median of $\sim \lambda/25$ with a top whisker at $\sim \lambda/10$. Considering the associated $z$-averaged power differences $\bar{\delta}$ with a median and a top whisker at $\sim 15\,\%$ and $\sim 40\,\%$, respectively, it becomes clear that even for large deviations between synthetic and reconstructed intensity distributions up to $\bar{\delta} \cong 40\,\%$ an accuracy for the retrieved phase of $\bar{\Phi}_{\text{RMSE}} \lesssim \lambda/15$ can be expected. Moreover, the outliers are mainly due to high noise level ($<\unit[20]{dB}$). This reveals that the noise has a strong impact and the detector should be well operated in terms of exposure time and dynamical range. Please note that for true experimental data (Sec.~\ref{section:4}), $\bar{\delta}, \delta_{z_j}$ are available and, thus, using the Monte Carlo result with the proposed fitting function $\bar{\Phi}_{\text{RMSE}}\left(\bar{\delta}\right)$ the quality of the retrieved phase profile can be roughly estimated \cite{merx2020beam}. If the intensity error $\bar{\delta}$ is several factors larger than the median, for example $\bar{\delta} \gtrsim 100\,\%$, the phase retrieval result should be questioned and the circumstances of the measurement, particularly any noise present, should be reviewed.

\subsection{Virtual backpropagation}\label{section:3b}
Knowledge about the optical field in amplitude and phase including phase discontinuities, see Fig.~\ref{fig:phasemaske1}\,(h) and (k), provides a major advantage for laser engineers, cf.~Sec.~\ref{section:1}. It helps to uncover potential weak points within the laser system or the beam path which are not accessible, for example, for mechanical reasons. Radiation that is truncated by apertures or that is illuminating damaged optical components will carry this information. During propagation, however, their impact can often only be recognized indirectly through beam quality deterioration and diffraction effects \cite{martinez1993second}. Knowledge of the underlying optical design---in the simplest case distances between mirrors in free-space propagation---allows to identify the locations of damaged components or faulty alignments. To do this, the optical field must be virtually propagated back through the optical system, for example, by applying the angular spectrum method \cite{goodman2005introduction}, cf.~Eq.~(\ref{eq:masterprop}).

\begin{figure*} []
    \centering
    \includegraphics[width=0.81\textwidth]{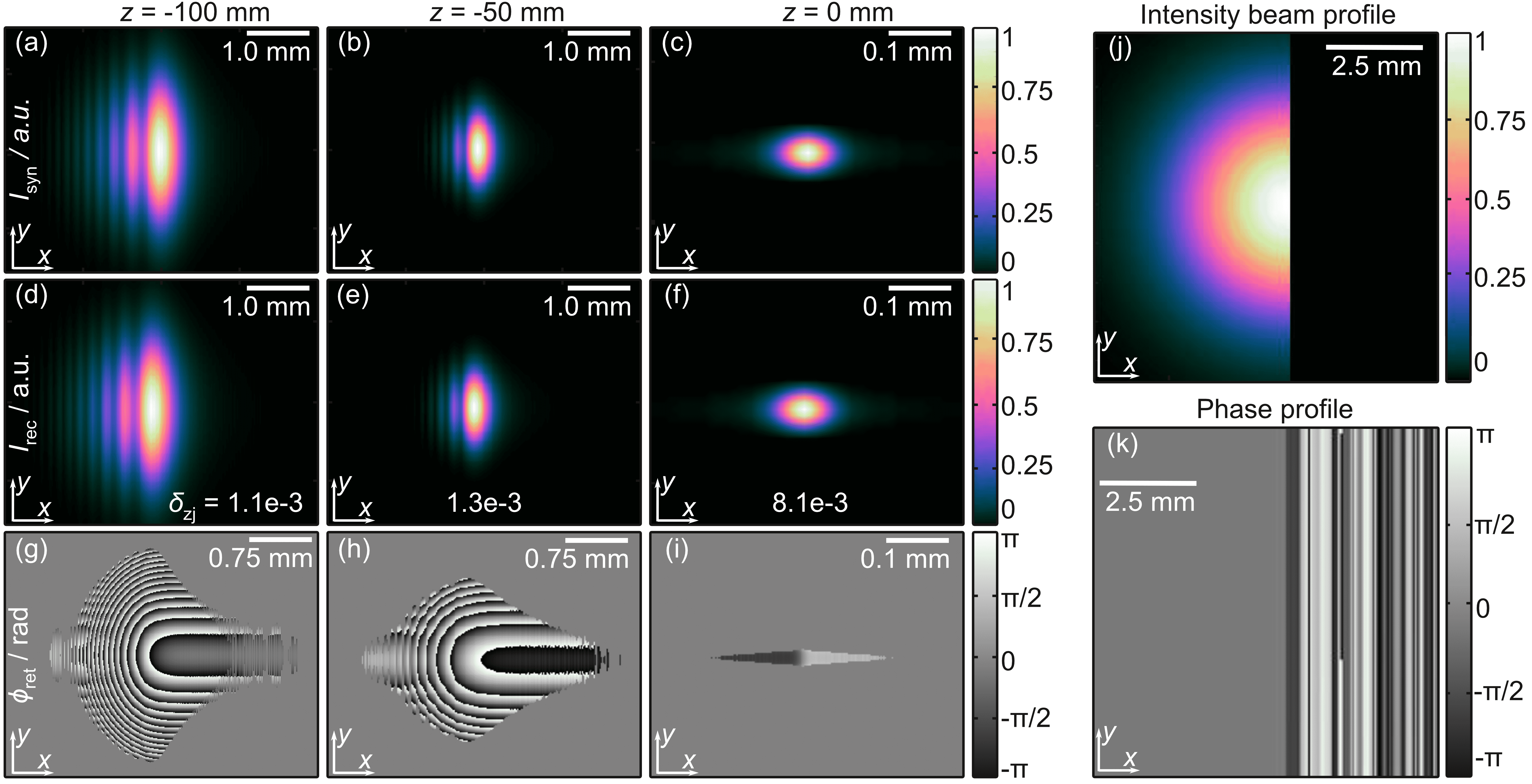}
    \caption {Details of the synthetic caustic of a truncated Gaussian-like beam $I_{\text{syn}}\left(x,y;z_j\right)$ at three different $z$-positions (a) -- (c). Corresponding reconstructed intensity $I_{\text{rec}}\left(x,y;z_j\right)$ and phase distribution $\phi_{\text{ret}}\left(x,y;z_j\right)$ (d) -- (i) with $\delta_{z_j}$-parameters well below $1\,\%$. Result of the virtual back propagation to the lens' back focal plane in intensity (j) and phase (k), respectively, with a clear identification of the truncating aperture.}
    \label{fig:sys_back}
 \end{figure*}

This ability is demonstrated by means of a selected synthetic example. Figure \ref{fig:sys_back} shows the propagation characteristics of a Gaussian-like beam which is obviously aberrated, see subfigures (a) -- (c) and corresponding $z$-positions after focusing with a lens.
Only from these intensity signals, it is hard to predict which errors in the beam path lead to these characteristic diffraction effects. As we know, the corresponding optical field from applying the phase retrieval algorithm (d) -- (i) with highest fidelity ($\delta_{z_j} < 1\,\%$), we can reconstruct the raw beam from propagating the optical field back into the lens' back focal plane. Here, the reason for focus quality degradation becomes clearly visible. The Gaussian raw beam was truncated by an aperture, see Fig.~\ref{fig:sys_back}\,(j). We also see that only amplitude disturbances lead to the present propagation behavior. The phase of the raw beam is flat, see Fig.~\ref{fig:sys_back}\,(k). Please note that the strong phase fluctuations in the right-hand side of Fig.~\ref{fig:sys_back}\,(k) are due to numerical noise at points of lowest intensities and are, thus, not relevant.

\section{Experimental results}\label{section:4}
We have discussed the capabilities and physical limitations of the iterative phase retrieval approach and will now apply the method to four selected cases. In Sec.~\ref{section:4}\,\ref{section4a} we will look at two examples and the ability to detect weaknesses in the optical path by virtual backpropagation. An Innoslab amplifier-like radiation is analysed in Sec.~\ref{section:4}\,\ref{section4c}. Finally, we discuss the field reconstruction from radiation generated by a thin-disk multipass amplifier, see Sec.~\ref{section:4}\,\ref{sec:section4d}.

\subsection{Virtual backpropagation of experimental data}\label{section4a}
\begin{figure}[t]
    \centering
    \includegraphics[width=0.47\textwidth]{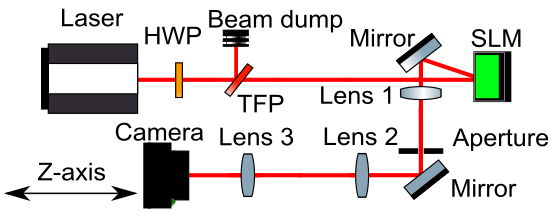}
    \caption {Experimental setup. Thin film polarizer (TFP), with half wave-plate (HWP) to attenuate the beam power. Liquid-crystal-on-silicon-based spatial light modulator (SLM). Imaging setup ($4f$-like) consisting of Lens 1 and 2. Lens 3 to generate the caustic recorded by camera with varying $z$-position.}
    \label{fig:experiment Setup_1}
\end{figure}

For experiments in this and the following section, the optical setup depicted in Fig.~\ref{fig:experiment Setup_1} is used. A low-power fiber laser of $P=\unit[24]{mW}$ average power, a beam diameter of $d_0=\unit[4]{mm}$, a pulse duration of ${\tau_{p}}=\unit[120]{ps}$, operating at  ${\lambda}=\unit[1030]{nm}$ serves as light source. A half-wave plate in combination with a thin-film polarizer (TFP) allows to select suitable intensities for our ISO-conform caustic measurements with camera IDS UI-3370CP-NIR-GL mounted on a $z$-axis stage. The liquid-crystal-based spatial light modulator (SLM, Hamamatsu X15223 series) and the following $4f$-setup can be neglected in this subsection. Here, the SLM acts as mirror only which is imaged onto the back focal plane of Lens 3 ($f=\unit[300]{mm}$). 

\begin{figure}[t]
    \centering
    \includegraphics[width=0.47\textwidth] {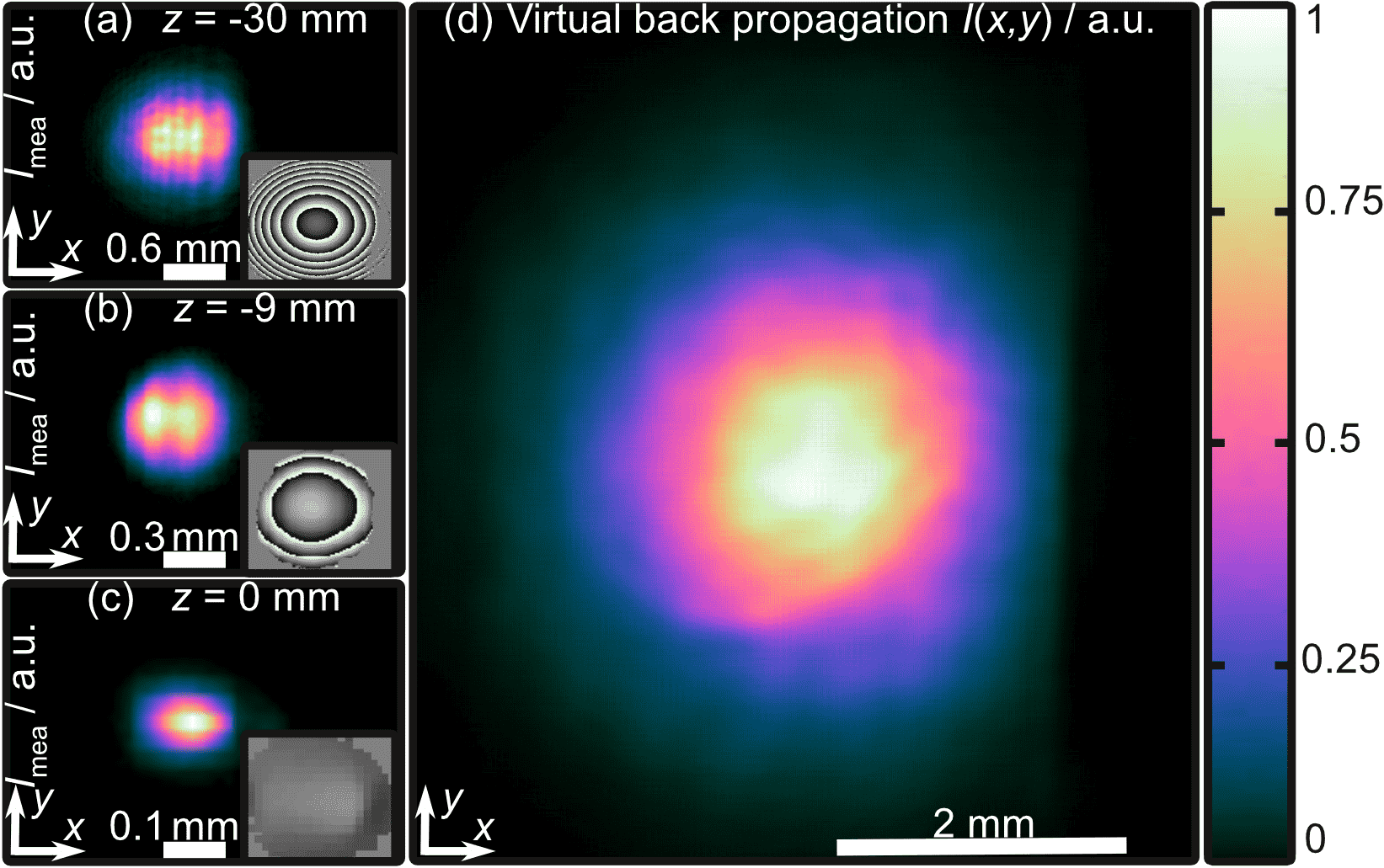}
    \caption {Virtual back propagation of an aperture-truncated Gaussian-like beam. Details of the ISO-standard conform caustic measurement  $I_{\textit{j}}\left(x,y;z_j\right)$ (a) -- (c) with  retrieved phase profiles $\phi_{\text{ret}}\left(x,y;z_j\right)$. The back-propagated intensity signal $I{\left(x,y\right)}$ reveals position of a single-side aperture (d).}
  
    \label{fig:aperture}
\end{figure}

In our experiment we mimic different amplitude distortions to the raw beam. The first one is similar to the virtual experiment discussed in Sec.~\ref{section:3}\,\ref{section:3b} where a straight aperture is blocking a portion ($\sim 2\,\%$ of power) of the raw beam (near-field). The impact to the far-field is visible in the recorded caustic, see Fig.~\ref{fig:aperture}\;(a) -- (c), with the retrieved phase distribution in the insets. In the focus (far-field) at $z = 0$, an elliptical focus shape (intensity) and a non-flat phase distribution ($\sim\lambda/3$ PV difference) is reconstructed indicating a disturbed beam propagation. The virtual propagation to the plane of Lens 3 reveals details of the aperture and the $z$-position in the beam path, see right hand side of Fig.~\ref{fig:aperture}\;(d).

\begin{figure}[t]
    \centering
    \includegraphics[width=0.47\textwidth] {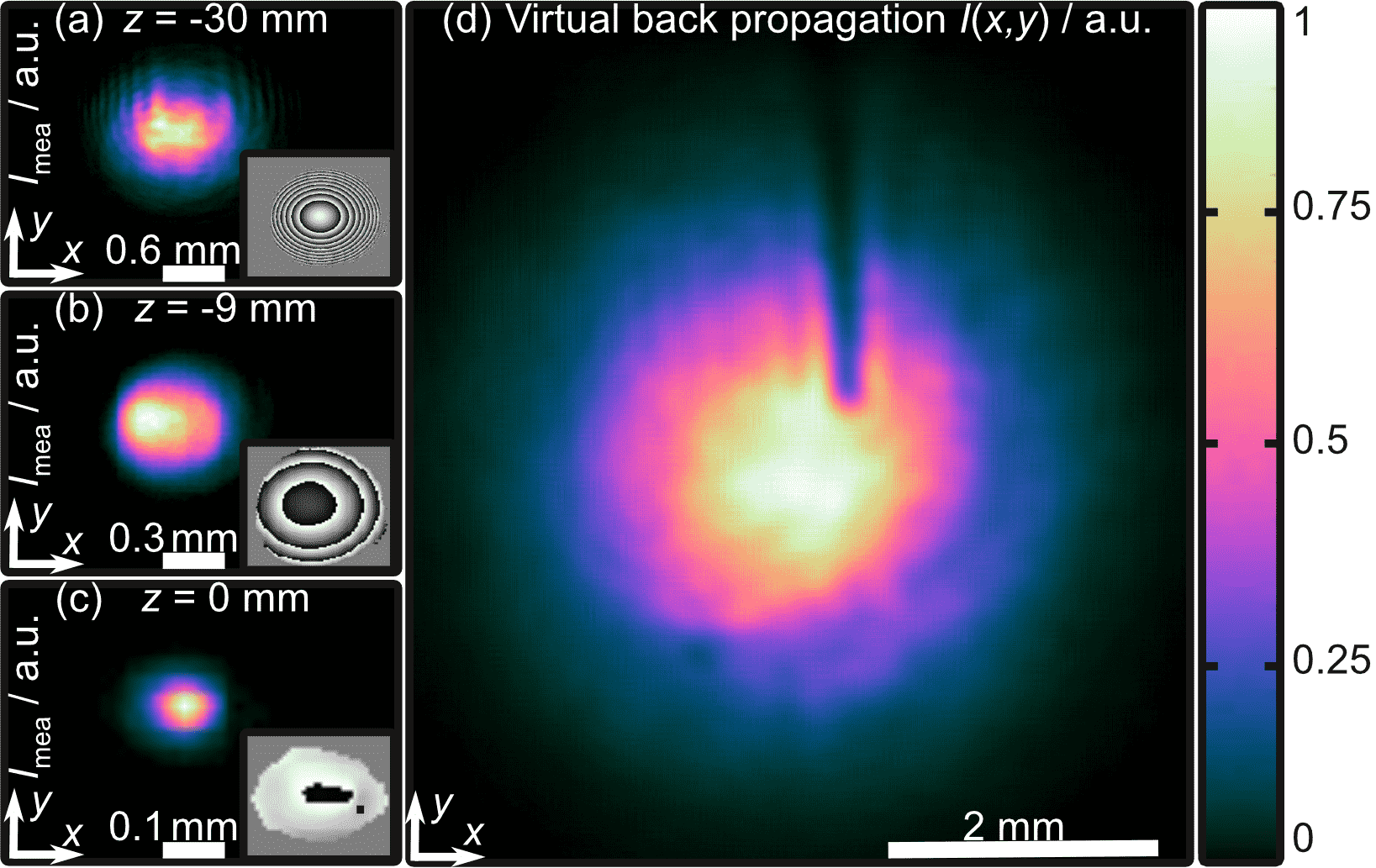}
    \caption {Virtual back propagation of an obscured Gaussian-like beam. Details of the ISO-standard conform caustic measurement $I_{\textit{j}}\left(x,y;z_j\right)$ (a) -- (c), with retrieved phase profiles $\phi_{\text{ret}}\left(x,y;z_j\right)$. The back-propagated intensity signal $I{\left(x,y\right)}$ reveals position and type of the obscuration---in this particular case a human hair of $\sim \unit[50]{\upmu m}$ width on the focusing lens.}
    \label{fig:hair_on_lens}
\end{figure}

With a second experiment we would like to demonstrate how effective the method is in identifying even smallest weak points in the beam path, see Fig.~\ref{fig:hair_on_lens}. We place a micrometer-scaled obstacle on an optical component (Lens 3). The associated power loss is barely detectable ($\ll 1\,\%$). However, considering the three details of our ISO-standard conform caustic measurement at $z = \unit[\left(-30, -9, 0\right)]{mm}$ diffraction effects are apparent from which position and shape of the obstacle can be identified, see Fig.~\ref{fig:hair_on_lens}\;(d).
In the present case a human hair of $\sim \unit[50]{\upmu m}$ width was placed on the focusing lens. For estimating smallest detectable amplitude modulations, the Rayleigh criterion is used. With a given wavelength $\lambda \approx \unit[1]{\upmu m}$ and numerical aperture of the focusing unit $\text{NA} \approx 0.03$ the resolution limit is $\approx \unit[20]{\upmu m}$. Smaller obscurations can be detected by increasing the employed NA.

\subsection{Innoslab-like laser radiation}\label{section4c}
Today's power and energy records in industry-grade ultrashort pulsed lasers are often generated by Innoslab amplifiers \cite{sutter2020next}. Here, longitudinally pumped slab-shaped laser media generate multi-millijoule and multi-hundred-watt in pulse energy and average power, respectively, at the one-picosecond-level with excellent beam quality $M^2_{\text{eff}} = \sqrt{M^2_xM^2_y} \lesssim 1.3$ \cite{russbueldt2014innoslab}. Although this beam quality---determined in accordance with the ISO standard---can actually be described as diffraction-limited, it exhibits characteristic features. A ``sinc''-like far-field pattern (usually filtered, see Ref.~\cite{russbueldt2014innoslab}) caused by diffraction at a gain aperture within the slab results in a distinct modulation of the intensity in the near-field \cite{russbueldt2014innoslab}. Although the overlap integral to the optical field of a fundamental Gaussian beam will be very close to $1$, the mentioned intensity modulations might have a strong impact for micro-machining \cite{russbueldt2014innoslab}. For this reason, we use the setup shown in Fig.~\ref{fig:experiment Setup_1} and generate an Innoslab-like laser beam which is analysed using the iterative phase retrieval. Here, the SLM is displaying a phase mask where the optical field's complex amplitude of an Innoslab-like laser is coded into a phase-only distribution \cite{arrizon2007pixelated}. The aperture after the SLM's first $2f$-setup, cf.~Fig.~\ref{fig:experiment Setup_1}, is used to block light in unwanted diffraction orders.

\begin{figure*}[]
   \centering
   \includegraphics[width=.8\textwidth]{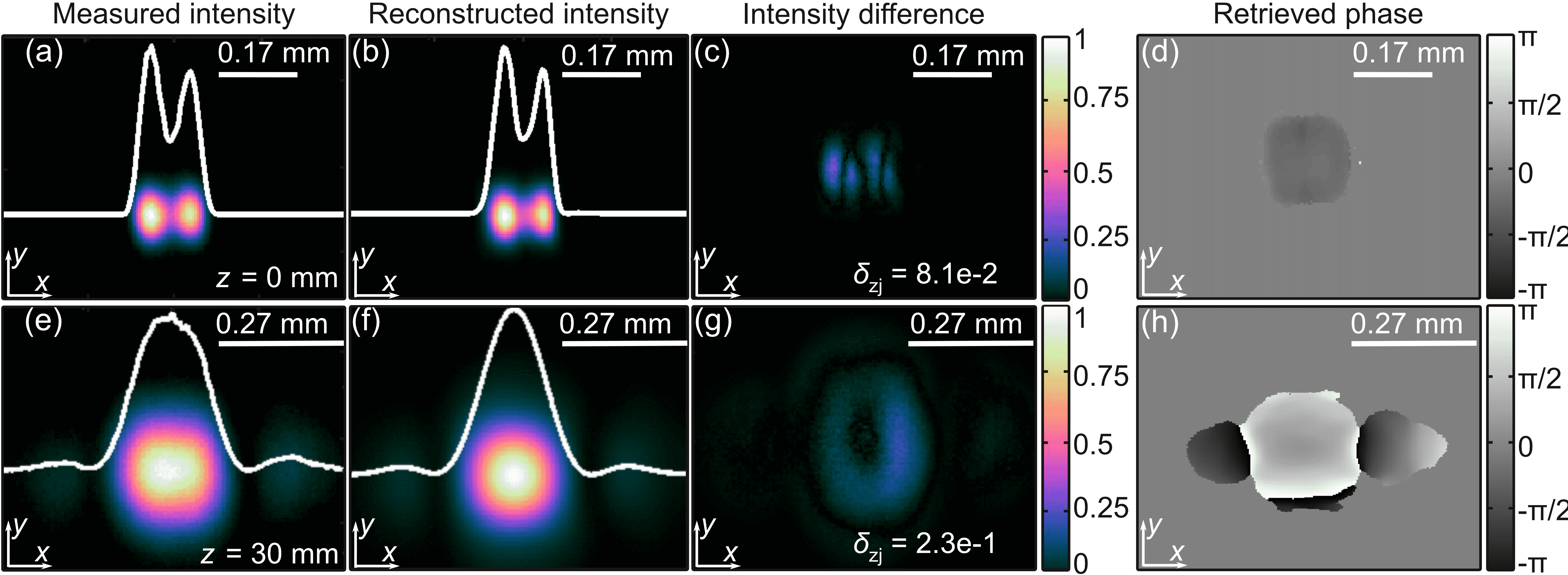}
   \caption{Innoslab-like laser phase retrieval result with measured near (a) and sinc-like far-field (e) (usually filtered \cite{russbueldt2014innoslab}) at $z=0$ and $z=\unit[30]{mm}$, respectively. The corresponding reconstruction (b) and (f) as well as intensity differences (c) and (g), with $\delta_{z_j} \lesssim 23\,\%$. Continuous near-field phase distribution (d) and non-continuous phase in the far-field (h), see $\uppi$-phase jumps (after subtracting defocus).}
   \label{fig:innoslab}
\end{figure*}

Results of this beam shaping are shown in Fig.~\ref{fig:innoslab} where (a) shows the modulated near-field at $z = 0$ and (e) the sinc-like far-field intensity distribution at $z = \unit[30]{mm}$.
This defocus is already sufficient for the beam to form its characteristic far-field distribution with well-known side lobes \cite{russbueldt2014innoslab}. Although the power in these side lobes amounts to only $3\,\%$ and the beam propagation ratio is still very well diffraction limited $M^2_{\text{eff}} \lesssim 1.3$, a strong intensity modulation is visible in the focus. The fact that the beam quality is nevertheless on this high level is revealed by the phase retrieval. The phase distribution at $z= 0$, see Fig.~\ref{fig:innoslab}\,(d) is almost plane (PV difference of $\phi\left(x,y\right) < \lambda/10$). Please note, that integrated intensity differences, see subfigures (c) and (g), amount to $\delta_{z_j} \lesssim 23\,\%$, cf.~Eq.~(\ref{eq:relative error}). We, therefore, expect deviations from the actual phase distribution to be better than $\bar{\Phi}_{\text{RMSE}} < \lambda/20$, see tolerancing in Sec.~\ref{section:3}\,\ref{section:montecarlo}.

In our first virtual experiment, cf.~Fig.~\ref{fig:phasemaske1}, we have emphasized the ability of our phase retrieval to retrieve non-continuous phase distributions. The phase profile of the present raw beam, generated with the SLM exhibits $\uppi$-phase jumps, too. These are located just between the side-lobes and the on-axis main lobe which is typical for sinc-like optical fields. This ability has now also been proven in experiments, see two vertical $\uppi$-phase jumps in Fig.~\ref{fig:innoslab}\,(h).

\subsection{High-energy ultrafast lasers}\label{sec:section4d}
A promising approach to scale ultrashort laser pulses to the $\sim\unit[100]{mJ}$ pulse energy regime with multiple-kilohertz repetition rates is the amplification with a thin-disk multipass cell (MPC) \cite{dominik2022thin}. Due to the thermal and mechanical stability as well as the preservation of the flexibility of the seed laser, this amplifier architecture is particularly relevant for industrial solutions \cite{dominik2022thin}. Therefore, we apply the phase retrieval approach to the radiation from a MPC amplifier to investigate the quality of its optical field.

\begin{figure}[t]
    \centering
    \includegraphics[width=0.47\textwidth]{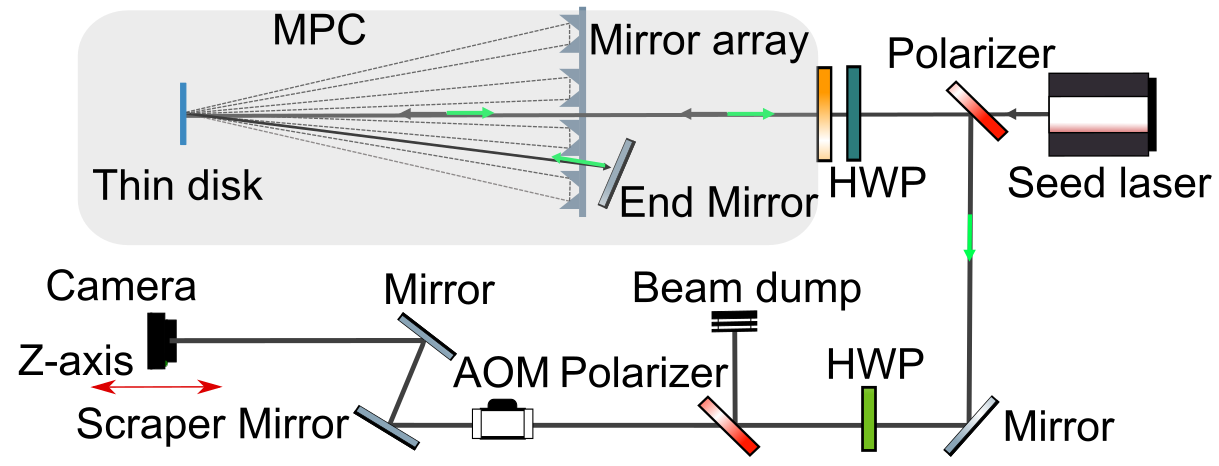}
    \caption {Experimental setup with a thin-disk multi-pass amplifier, containing thin disk, mirror array and end mirror. Half-wave plate, polarizer, acousto-optic modulator (AOM), and scraper mirror for attenuating extreme laser pulses.}
    \label{fig:Experimental setup}
\end{figure}

To combine mentioned multi-millijoule pulse energies with sub-picosecond pulse durations, the MPC is seeded with an Innoslab amplifier. This combination will be realized in a future industrial product---the TRUMPF TruMicro Series 9000. Here, experiments were carried out on an adapted setup of this architecture enabling kilowatt-level output powers but with longer pulse durations. The MPC amplifier running in double-pass mode and seeded by an adapted TruMicro 6000 Series laser operating at $\lambda = \unit[1030]{nm}$ provides $\unit[120]{ps}$-pulses with a repetition of $\unit[6]{kHz}$, see setup in Fig.~\ref{fig:Experimental setup}. After amplification the pulses exhibit $\unit[50]{mJ}$ energy resulting in $\unit[280]{W}$ average power.

In a next step these extreme pulses need to be prepared for the laser beam characterization. Thus, they need to be attenuated carefully without altering the spatio-temporal pulse properties. The combination of half-wave plate (HWP) and polarizer is disposing of already $\unit[23]{dB}$ of pulse energy and average power, respectively. A further attenuation of $\unit[13]{dB}$ is achieved with the acousto-optic modulator (AOM) and the scraper mirror where the first diffraction order is selected with a well-defined diffraction efficiency (without pulse picking). Finally, neutral density filters are further absorbing $\unit[2]{dB}$ of average and peak power before the beam illuminates a commercial tool for laser beam characterization (Metrolux LPM200-YAG). Here, a focusing unit with $f = \unit[300]{mm}$ is generating a waist with $z_\text{R} \approx \unit[5]{mm}$ which is sampled by a set of $20$ $I_j\left(x,y\right)$-measurements, cf.~Sec.~\ref{section:2}. When using commercial tools for recording the caustics $I_j\left(x,y\right)$ and for applying the iterative phase retrieval, please note the following. The spatial assignment of the intensity profiles must be carried out along a global optical axis defined, for example, by the axis stage moving the camera. In some cases, however, commercial tools for caustics measurements are defining a region-of-interest with a local coordinate system defined, for example, by the intensity's first-order-moment $I_j\left(x,y\right) \rightarrow I_j\left(x_j',y_j'\right)$. However, in general, this local optical axis is not identical with the global one especially when general astigmatic beams are investigated \cite{wielandy2007implications}.

The iterative phase-retrieval algorithm will also fail, if the optical field of the laser under test will change during the recording. A typical caustics measurement process requires a few minutes. Within this period the present optical field as well as positioning and pointing stability should stay constant, see, for example Monte Carlo simulations in Sec.~\ref{section:3}\,\ref{section:montecarlo} with variation of detector shift. Considering the radiation under test, we
investigated the beam's first-order moment using the Metrolux LPM200-YAG within a period of $\unit[2]{h}$. Here, the beam's pointing was measured to $\lesssim \unit[2]{\upmu rad}$. In cases where real-time analysis of optical fields are required, we refer to solutions as proposed in Ref.~\cite{scaggs2012real} where multiple defocused intensity signals can be recorded simultaneously. Real-time laser beam characterization (including phase retrieval) is particularly beneficial when high-power amplifiers are considered, and the impact of the thermal loads need to be investigated. In general, if the metrology's recording time cannot sample the temporal fluctuations of the beam under test, this would be directly visible in the quality of the amplitude reconstruction $\bar{\delta}$, cf.~Sec.~\ref{section:3}\,\ref{section:montecarlo}.

\begin{figure*}[]
   \centering
   \includegraphics[width=0.8\textwidth]{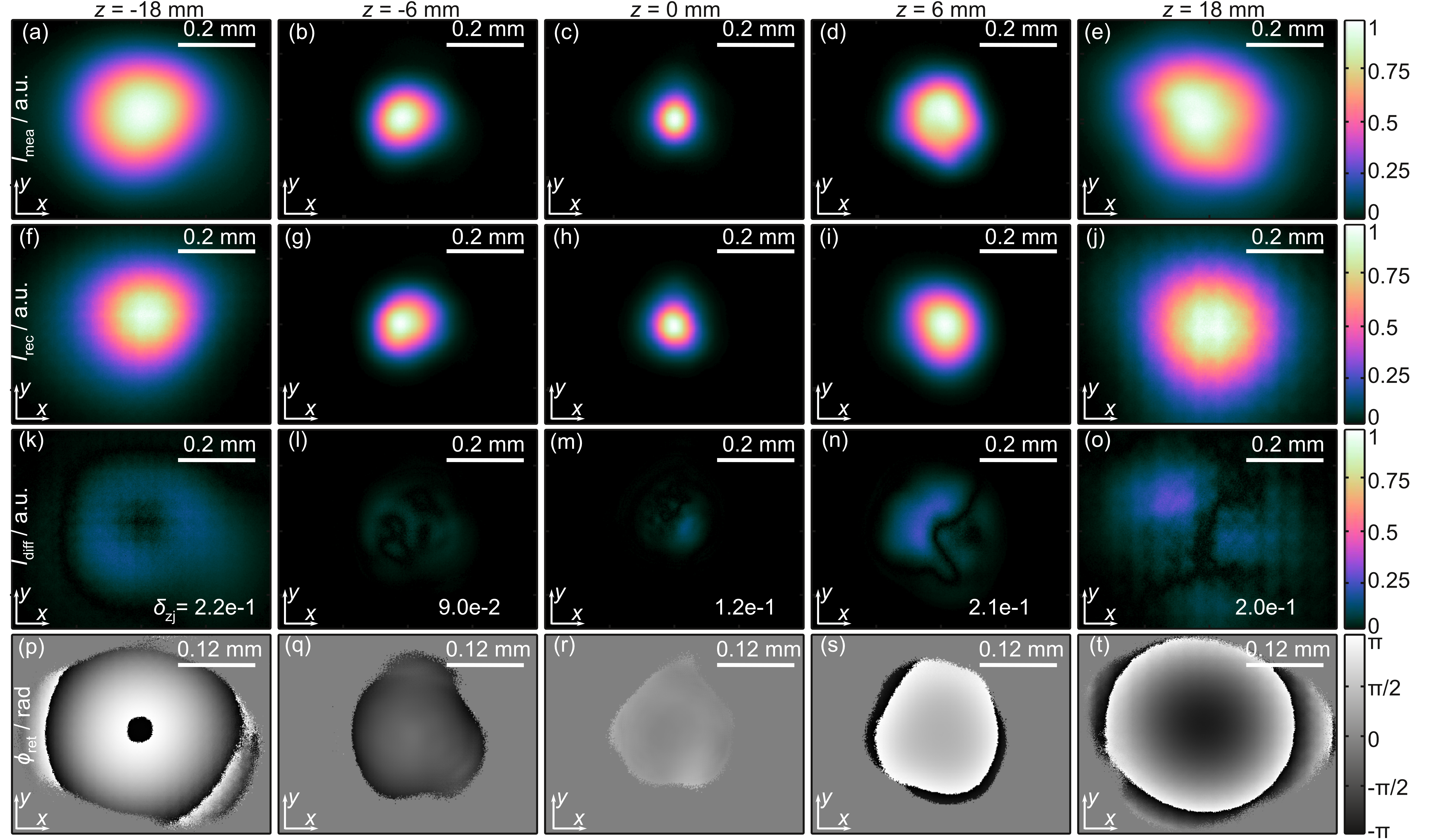}
   \caption{Phase retrieval result from analysing $\unit[50]{mJ}$-energy ultrashort pulses emerging from a multipass cell laser amplifier. Details of the measured caustic (a) -- (e) with reconstructed intensity distributions (f) -- (j) and corresponding intensity differences (k) -- (o) with $\delta_{z_j} \leq 0.22$. Resulting phase profiles (p) -- (t) proving excellent beam quality with a nearly plane phase in the focus at $z=0$ (r).}
   \label{fig:Exp_beam_1.2}
\end{figure*}

Figure \ref{fig:Exp_beam_1.2} provides the results of the phase retrieval analysis applied to the radiation emerging from our high-energy ultrafast laser.
Again, we compare measured (a) -- (e) with reconstructed intensity signals (f) -- (j) at different $z$ positions (selected details of the entire caustic measurement). The  quality of these reconstructions is available from the $\delta_{z_j}$-parameter, cf.~Eq.~(\ref{eq:relative error}), being less than $0.22$ at any considered propagation step, see subfigures (k) -- (o). Considering our tolerance study, cf.~Fig.~\ref{fig:power vs rmse}, we expect deviations of the retrieved phase distribution $\phi_{\text{ret}}\left(x,y\right)$ to be smaller than $\lambda/25$ in terms of $z$-averaged root-mean-square error, cf.~Eq.~(\ref{eq:rmsebar}). In the last row of Fig.~\ref{fig:Exp_beam_1.2} the retrieved phase distributions are depicted. The defocus is not removed in any of the subfigures (p) -- (t) and vanishes only in the focus, see (r). Here, a phase profile is at hand which is almost plane---without phase jumps or singularities. The PV difference amounts to $<\lambda/6$ ensuring a large overlap integral to a fundamental Gaussian beam ($>\unit[96]{\%})$. The Gaussian-like amplitude profile (focus circularity $>87\,\%$) in combination with the plane phase corresponds to a diffraction-limited beam quality of $M^2_{\text{eff}} \leq 1.2$. With the optical field uncovered in this way, optical tools can now be designed for sophisticated micro-material processing on macroscopic surfaces or volumes \cite{jenne2020facilitated, ranke2022high}.

\section{Conclusion}
To conclude, we have combined the metrology for determining the beam quality from intensity profiles according to the ISO standard with a method for retrieving phase distributions. After estimating the limits of the metrology with a Monte Carlo simulation, it was applied to various cases relevant for materials processing. Using standard-conform measuring devices for laser beam characterization, phase reconstructions with an accuracy down to $\lambda/25$ (RMSE) can be expected---including cases where non-continuous profiles are present. We have demonstrated the potential of our approach to uncover possible weak spots in the beam path by virtually back-propagating the reconstructed optical field. Finally, $\unit[50]{mJ}$-energy pulses from an ultrafast laser source have been spatially analysed. In addition to the global parameters for laser beam quality, spatial phase distributions are obtained. The equipment required is limited to ISO-standard conform tools that are usually available in laser laboratories.

\section*{Funding}
Supported by the Free State of Thuringia and the European Social Fund Plus (2022FGR0002) and the German Federal Ministry of Education and Research (BMBF) (RUBIN-UKPino 03RU2R032F).

\section*{Disclosures}
JW, SSR, DF: TRUMPF Laser- und Systemtechnik GmbH (E); SB, BD, MS, DB: TRUMPF Laser GmbH (E).

\section*{Data Availability}
Data underlying the results presented in this paper are not publicly available at this time but may be obtained from the authors upon reasonable request.

\bibliography{Lib1}
\bigskip
\noindent
\bibliographyfullrefs{Lib1}

\end{document}